\documentclass[prl,aps,twocolumn,showpacs]{revtex4} 
\usepackage{graphicx} 
\usepackage{dcolumn}

\begin{document} 
 
\title{ 
Origin of spontaneous broken mirror symmetry of  vortex lattices in Nb
} 
\author{H. M. Adachi} 
\affiliation{Department of Physics, Okayama University, 
Okayama 700-8530, Japan}  
\author{M. Ishikawa} 
\affiliation{Department of Physics, Okayama University, 
Okayama 700-8530, Japan}  
\author{T. Hirano} 
\affiliation{Department of Physics, Okayama University, 
Okayama 700-8530, Japan} 
\author{M.  Ichioka} 
\affiliation{Department of Physics, Okayama University, 
Okayama 700-8530, Japan} 
\author{K.  Machida} 
\affiliation{Department of Physics, Okayama University, 
Okayama 700-8530, Japan} 
\date{\today}

\begin{abstract} 
Combining the microscopic Eilenberger theory with the first
principles band calculation, we investigate
the stable flux line lattice (FLL) for a field applied to the four-fold axis;
$H\parallel [001]$ in cubic Nb.
The observed FLL transformation along $H_{c2}$ is
almost perfectly explained without adjustable parameter, including
the tilted square, scalene triangle with broken mirror symmetry, and
isosceles triangle lattices upon increasing $T$. We construct a minimum Fermi 
surface model  to understand those morphologies, in particular
the stability of the scalene triangle lattice attributed to the lack of the mirror symmetry
about the Fermi velocity maximum direction in k-space.
\end{abstract}

 

\maketitle 

A series of elemental metal superconductors, Nb, V, Al, $\cdots$
have a long history since its discovery of  Hg in 1911.
Those are now regarded as a ``conventional'' superconductor where
in fact  the energy gap is isotropic and electron-phonon mechanism is known to work well.
Thus it seems that no mystery remains there and every superconducting aspect
is  well understood and controlled since Nb is known to be wide practical applications
from Nb-SQUID magnetometers to superconducting cavities for particle accelerators\cite{nature}.
However, recent small angle neutron scattering (SANS)
experiments on Nb discover  plethora of the mysterious
vortex morphology\cite{laver1,laver2,muhlbauer}, 
requiring us  a consistent understanding of flux line lattice 
symmetry changes upon varying the applied field direction continuously, that should satisfy
a global constraint imposed by the hairy ball theorem\cite{hairy},
namely any local change of FLL  symmetry must be consistent with
the change viewed globally.

Among them recent finding by Laver {\it et al.}~\cite{laver1}  is particularly intriguing:
They find a novel FLL whose half unit cell is a scalene triangle
by SANS experiment on Nb under a field applied along $H\parallel [001]$
direction. By varying $H$ and $T$ they have succeeded in constructing 
the vortex lattice phase diagram (see Fig.1(b)).
Along the $H_{c2}$ line from low $T$ and high $H$ the
square lattice tilted by $10.9^{\circ}$ from the [100] direction changes into
a scalene lattice followed by an isosceles lattice that is stabilized at higher $T$.
Although the tilted square lattice has been known for a quite long time
experimentally without any theoretical understanding\cite{weber},
the discovery of the scalene lattice is new and intriguing since the scalene lattice breaks the 
fundamental mirror symmetry spontaneously in cubic crystalline Nb.
Notice that among various superconductors, conventional and unconventional,
there is no known case to exhibit either scalene triangle or tilted square
FLL under the four-fold symmetric field configuration, namely,
tetragonal crystals under $H\parallel [001]$ such as CeCoIn$_5$~\cite{bianch},
Sr$_2$RuO$_4$~\cite{riseman}, La$_{1.83}$Sr$_{0.17}$O$_4$~\cite{gilardi}, 
ErNi$_2$B$_2$C~\cite{morten1}, YNi$_2$B$_2$C~\cite{morten2},
and TmNi$_2$B$_2$C~\cite{debeer}.

According to Takanaka~\cite{takanaka} who is a pioneer in this field of theoretical FFL
morphology studies, the free energy of FLL contains the
 so-called phasing energy that depends on the relative
orientation between the FLL and underlying crystal lattice.
This implies that the  observed tilted  square lattice may
be stabilized by this phasing energy. This is expanded in terms of the higher order
harmonics of the four-fold symmetry ($\theta$ is the angle from the [100] direction), 
$F_{phasing}= \Sigma_{n=1,2,\cdots}A_{4n}\cos4n\theta$
whose minima may give rise to the tilting angle $\theta=10.9^{\circ}$
from the [100] direction, meaning that at least 
two or more higher harmonics are simultaneously non-vanishing.

Here the purposes of this paper are two-fold:
The first principles band calculation based on density functional theory (DFT) 
yields precise information on the 
electronic properties of the normal state in general.
By combining it with the microscopic Eilenberger theory,
we can establish a truly first principles framework for a
superconductor, in particularly for vortex state under an applied field
without any adjustable parameters.
In order to benchmark it we apply this to explain those
 intriguing vortex morphologies, which will turn out that this 
 framework is remarkably successful. 
 
 Thus the second
 purpose is to understand the physical origin for the scalene 
 FLL structure by constructing a minimum model Hamiltonian to
 describe the FLL transformation mentioned above.
 The mirror symmetry breaking with the scalene FLL is 
 attributed to the special character of the Nb electronic band structure.
 This study points to the fact that our combined  framework of DFT and the
 Eilenberger theory paves the way to truly first principles theory for superconductors 
 properties under a field without adjustable parameter.
 Previously  several attempts have been done to perform this program, but those 
 are limited to either discussions on $H_{c2}$\cite{kita} or resorting to approximate
 solutions  for Eilenberger equation\cite{nagai}.
 
 In order that our calculations are tractable within a reasonable computational
 time frame, we limit our discussion along $H_{c2}$, that reduces significantly
 the computational time because we must minimize the free energy
 in the multi-dimensional parameter space spanned by the FLL unit cell shape and
 its orientation relative to the underlying crystal axes.

 The quasiclassical Eilenberger theory\cite{eilenberger,ichioka,nakai}  is quantitatively valid when 
$\xi \gg 1/k_{\rm F}$ ($k_{\rm F}$ is the Fermi wave number 
and $\xi$ is the superconducting coherence length).
The quasiclassical Green's functions 
$g( \omega_n, {\bf k},{\bf r})$,  
$f( \omega_n, {\bf k},{\bf r})$, and  
$f^\dagger( \omega_n, {\bf k},{\bf r})$  
are calculated by the Eilenberger equation 
\begin{eqnarray} && 
\left\{ \omega_n  
+{\bf v}_F \cdot\left(\nabla+{\rm i}{\bf A} \right)\right\} f( \omega_n, {\bf k},{\bf r}) 
=\Delta g( \omega_n, {\bf k},{\bf r}),  
\nonumber  
\\ &&  
\left\{ \omega_n  
-{\bf v}_F  \cdot\left( \nabla-{\rm i}{\bf A} \right)\right\} f^\dagger( \omega_n, {\bf k},{\bf r}) 
=\Delta^\ast g( \omega_n, {\bf k},{\bf r}),  \quad  
\label{eq:Eil} 
\end{eqnarray}  
where $g=(1-ff^\dagger)^{1/2}$, ${\rm Re}$$g > 0$,  
$\bf A$ is the vector potential and ${\bf v}_F $ is the Fermi velocity. 
 At $t=T/T_c$, the self-consistent gap equation is given by
 \begin{eqnarray}
 \Delta\ln t= t \sum_{\omega_n}\left(\langle f(\omega_n, {\bf k},{\bf r})\rangle_{\bf k}
 -{\Delta\over|\omega_n|} \right), 
\end{eqnarray}  
where $\langle\cdots\rangle_{\bf k}=\int d\theta D(\theta, \epsilon_F) (\cdots)$
with $D(\theta, \epsilon_F)$ being the angle-resolved density of states (AR-DOS) at the Fermi surface.

 Along $H_{c2}$, we linearize the Eilenberger equation as 
\begin{eqnarray}
[\omega_n+{\bf v}_F \cdot(\nabla+{\rm i}{\bf A})] f( \omega_n, {\bf k},{\bf r}) 
=\Delta.
\end{eqnarray}
By expanding  $\Delta$ and 
$f(\omega_n, {\bf k},{\bf r})$ in terms of the Landau-Bloch function 
$\psi_{N, {\bf q}}({\bf r})$ whose $N$-th coefficients are $\Delta_N$ and $f_N(\omega_n, {\bf k})$,
we can obtain the eigenvalue equation for $\Delta$, 
from which $H_{c2}$, $\Delta_N$ and $f_N(\omega_n, {\bf k})$ are evaluated\cite{kita2}.
Those eigenfunctions lead to the gap function $\Delta({\bf r})$ 
and the  quasiclassical Green's function $f( \omega_n, {\bf k},{\bf r})$.

In order to compare various FLL forms to find the most stable
vortex configuration near $H_{c2}$, we have to calculate the free energy $F$  given by 
 \begin{eqnarray}
F=\tilde{\kappa}^2\overline{(\nabla\times{\bf A})^2}+\overline{|\Delta|^2} \ln t 
+2t\sum_{\omega_n>0}\left[{\overline{|\Delta|^2}\over \omega_n}-\overline{\langle I\rangle}\right] 
 \end{eqnarray}
with $I=(g-1)[2\omega_n+{\bf v}_F\cdot(\nabla\ln(f/f^{\dagger})+2i{\bf A})]+
(f\Delta^{\ast}+\Delta f^{\dagger})$. 
Via the Abrikosov identity and expanding the normalization condition up to
the next order; $g\sim 1-ff^{\dagger}/2-(ff^{\dagger})^2/8$, 
we obtain the free energy valid near $H_{c2}$
as~\cite{nakai} 
 \begin{eqnarray} && 
 {F\over \tilde{\kappa}^2}=B^2-{(B-H_{c2})^2\over{\tilde{F}+1}}, 
\\ && 
\tilde{F}={t\over {4\tilde{\kappa}^2\bar{h}_s^2}}\sum_{\omega_n>0}
\overline{ 
 \langle{
ff^{\dagger}(f\Delta^{\ast}+\Delta f^{\dagger})}
\rangle_{\bf k}
}
-{\overline{h_s^2}\over \bar{h}_s^2}, 
\end{eqnarray}
where $\overline{(\cdots)}$ denotes the spatial average within a unit cell, 
 and $\bar{h}_s$ is the magnetic field induced by supercurrent.
$B=H+\bar{h}_s$ and $\tilde{\kappa}^2=7\zeta(3)\kappa^2/18$. Since we know that the vortex morphology
 is independent of GL parameter $\kappa$\cite{suzuki}, here we have used a large $\kappa$.
The free energy minimum corresponds to the minimum of $\tilde{F}$ that gives the stable 
vortex lattice configuration. 

\begin{figure}[tb] 
\includegraphics[width=8.5cm]{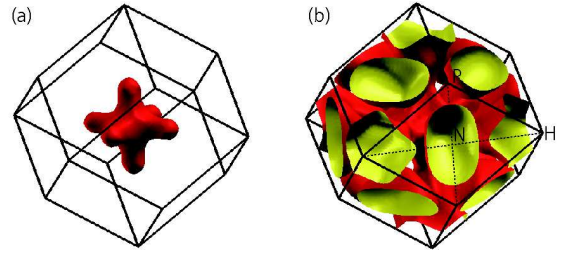} 
 \caption{ 
(Color online) Calculated Fermi surfaces (a) and (b) for Nb
in first Brillouin zone where the former (latter) is closed (extended).
} 
\label{fig:DFT} 
\end{figure}
\begin{figure}[tb] 
\includegraphics[width=8cm]{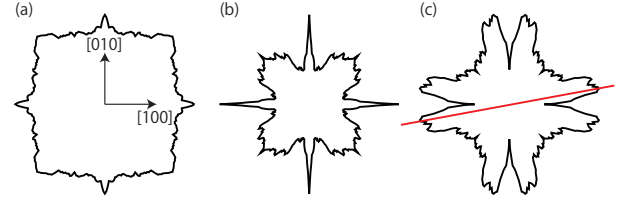} 
 \caption{ 
(Color online) 
The cross sections perpendicular to the field direction H$\parallel$
[001] for  the Fermi velocity anisotropy  $v_F(\theta)$ (a), AR-DOS $D(\theta, \epsilon_F)$ (b)
and the inverse of the AR-DOS (c) where the red line drawn on its  maximum direction
indicates the lack of the mirror symmetry about this axis tilted by 11$^\circ$ from [100].
} 
\label{fig:ARDOS} 
\end{figure}

The Fermi velocity ${\bf v}_F(\theta)$ and 
 AR-DOS $D(\theta, \epsilon_F)$ are evaluated by using tetrahedron method within DFT\cite{wien2k}.
 The first Brillouin zone is divided into 135$^3$ parallel pipes, each of which  is further subdivided
 into 6 pieces. 
 As shown in Fig. \ref{fig:DFT}, our DFT calculation reproduces the well-known 
two Fermi surfaces\cite{nb}. Then the band information
is used to yield  the Fermi velocity anisotropy $v_F(\theta)$ shown in Fig. \ref{fig:ARDOS}(a)
and the AR-DOS $D(\theta, \epsilon_F)$ in  Fig. \ref{fig:ARDOS}(b) where
we map the three dimensional objects onto the cross sections perpendicular 
to the field direction for the study of vortex states. Those quantities are 
used for evaluating the free energy. 
Here we also show the inverse of AR-DOS in 
Fig. \ref{fig:ARDOS}(c) for later purpose.

\begin{figure}[tb] 
\includegraphics[width=8.0cm]{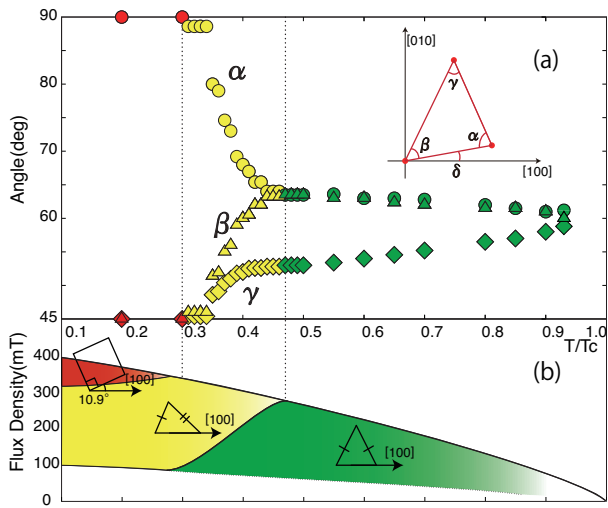} 
 \caption{ 
(Color online)  Calculated FLL transformation along $H_{c2}$ (a)
where the square lattice is tilted by $\delta$=14$^{\circ}$ from [100],
the scalene FLL is stabilized in intermediate $T$, and the isosceles triangle
in high $T$ is along [100].
Observed phase diagram in Nb (b). 
} 
\label{fig:L1} 
\end{figure} 

\begin{figure}[tb] 
\includegraphics[width=9cm]{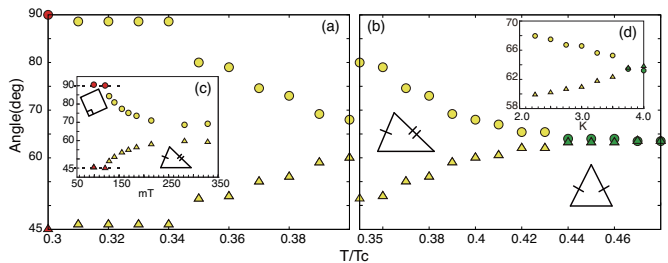} 
 \caption{ 
(Color online) Detailed FLL transformation of Fig.1
(a) low $T$ from tilted square to scalene triangle and (b) intermediate $T$
from the scalene to isosceles triangle. The insets (c) and (d) show the
experimental data\cite{laver1} for reference: (c) data at 2K as a function of $H$,
and (d) data at $H$=200mT as a function of $T$.
} 
\label{fig:L2} 
\end{figure} 

Figures \ref{fig:L1} and \ref{fig:L2} show the FLL transformation along $H_{c2}$ where
the unit cell shape characterized by the angles, $\alpha$, $\beta$ and $\gamma$ and the tilting angle 
$\delta$ from [100] are displayed in Fig. 3(a)
and compared with the observed phase diagram\cite{laver1} in $H$ vs $T$ for Nb
in Fig. 3 (b).
Note that the square lattice tilted by $\delta$=14$^{\circ}$ is stabilized at the low $T$ and high $H$
region. This tilting angle nearly coincides with the observed value 10.9$^{\circ}$
and their $H-T$ region also coincides with the observation.
Upon increasing $T$, the tilted square lattice changes into the
scalene lattice at $t=0.3$.  Then after gradual transformation, keeping the scalene nature
intact for a finite $H$ region. This scalene lattice that breaks the mirror symmetry of
FLL finally gives way to the isosceles triangle lattice at higher $T$.
This isosceles is orientated along the [100] direction ($\delta$=0$^{\circ}$) that coincides 
with the observation. This lock-in transition occurs smoothly.
Upon further increasing $T$, this isosceles triangle lattice tends to the
equilateral triangle toward $T_c$ smoothly.
The overall FFL transformation characterized by the unit cell shape and
orientation relative to the crystal lattice along $H_{c2}$  almost
perfectly reproduces the observation\cite{laver1,laver2}. It should be emphasized that
there is no adjustable parameter here and all information on the
electronic structures in the normal state, including the Fermi velocity anisotropy and
the AR-DOS is coming from the first principles DFT calculations.

In order to understand the physical origins of the remarkable 
success of the above Eilenberger theory combined with DFT,
we construct a minimum model for grasping the essential features
of those results. Namely (1) the tilted square at low $T$,
(2) the scalene triangle in the intermediate $T$ and 
(3) the isosceles triangle whose nearest neighbor is along [100]
at higher $T$.
We model the Fermi velocity anisotropy in the Fermi circle whose radius
is $k_F$, that is, ${\bf v}_F=\beta(\theta)v_F^0{\bf u}$.
$\beta(\theta)$  parametrizes the Fermi velocity anisotropy,
and ${\bf u}$ is the unit vector on the plane perpendicular to the
field direction. In this model, the AR-DOS
is given by $D(\theta, \epsilon_F)=1/\beta(\theta)v_F^0$.
Namely, the Fermi velocity anisotropy and AR-DOS are not
independent quantities as in DFT. 
We can expand as $\beta(\theta)=\Sigma_{n=0,1,2,\cdots}\beta_{4n}\cos(4n\theta)$.
Since it will be clear that $\beta_4$ only model cannot account the above
features (1)-(3), the minimum model consists of $\beta_4$ and $\beta_8$.
Namely, we consider the following model:
$${\bf v}_F(\theta)=v_F^0{\bf u}\beta(\theta)
=v_F^0{\bf u}(1+\beta_4\cos4\theta-\beta_8\cos8\theta).$$

We have performed extensive calculations based on this model, plugged it
into the Eilenberger equation above to find the conditions to reproduce
the above features. Then we have found the necessary condition for 
scalene triangle to appear as $\beta_8>\beta_4>0$
that also explains the features (1-3) simultaneously.

\begin{figure}[tb] 
\includegraphics[width=8.0cm]{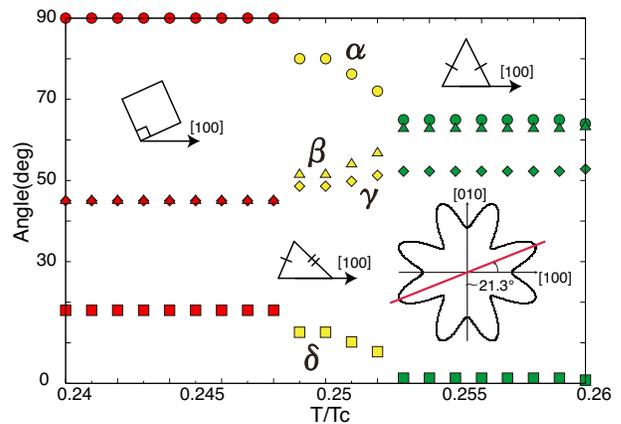} 
 \caption{ 
(Color online)  
FLL transformation based on the minimum model with $\beta_4=0.1$
and $\beta_8=0.3$. The low $T$ square lattice tilted with 18$^{\circ}$
from [100] axis transforms into the scalene FLL that is followed by 
the isosceles FLL at higher $T$ where the nearest neighbor orients
along [100] direction.
} 
\end{figure} 
 
Figure 5 shows an example for this case ($\beta_4=0.1$
and $\beta_8=0.3$). In the lowest $T$ region, the square lattice
appears with the tilting angle $\theta=18^{\circ}$ while
 the Fermi velocity maximum is at $\theta=21.3^{\circ}$.
 There is a rule that at lower $T$ the stable square lattice is oriented
 along either the AR-DOS maximum for isotropic gap case 
 or the nodal direction for d-wave gap case\cite{suzuki}.
 
 Then by increasing $H$, the tilted square deforms continuously
 into the scalene triangle FLL as seen from Fig. 5 in the middle
 $T$. This scalene lattice is locked-in the isosceles triangle lattice
 oriented along [100]. Those sequences of the FLL transformation
 are same as in DFT case above. It is now clear that
 the essential ingredient for the scalene triangle to appear is
 the lack of the mirror symmetry about the Fermi velocity maximum 
 direction as indicated in the inset of Fig. 5 and also note that
 in Fig. 2(c) the inverse of AR-DOS also shows this.
 This velocity maximum simultaneously explains why the square lattice is tilted.
 Namely the tilted square FLL and scalene FLL
 are intimately tied with each others.
 Note that while our $\beta_4$-$\beta_8$ minimum model
 explains the appearance of the scalene FLL, the stable $T$ region
 is far off the observation.

In order to understand the origin of the scalene FLL,
we display the deformation of this lattice in Fig. 6(a).
It is seen from it that the
deformation occurs so that one of the two diagonal axes
of the parallel piped unit cell continuously rotates,
keeping the other axis intact. This movement can be
interpreted as the scalene FLL seeks  stable configuration 
by adjusting the unit cell angle and its orientation by utilizing 
the possible freedoms allowed.
As a result of the broken mirror symmetry about the velocity maximum direction 
the scalene FLL could appear.
This situation is quite different from that in the $\beta_4$ only (the same for 
the $\beta_8$ only model) where the mirror symmetry is preserved in the velocity 
anisotropy along the maximum direction.
As shown in Fig. 6(b), the FLL deformation from the square to isosceles triangle proceeds 
as follows: The two diagonal axes of the parallel piped unit cell 
keep unchanged, always orthogonal with each other and the two angles of the isosceles 
continuously changes.

\begin{figure}[tb] 
\includegraphics[width=7.0cm]{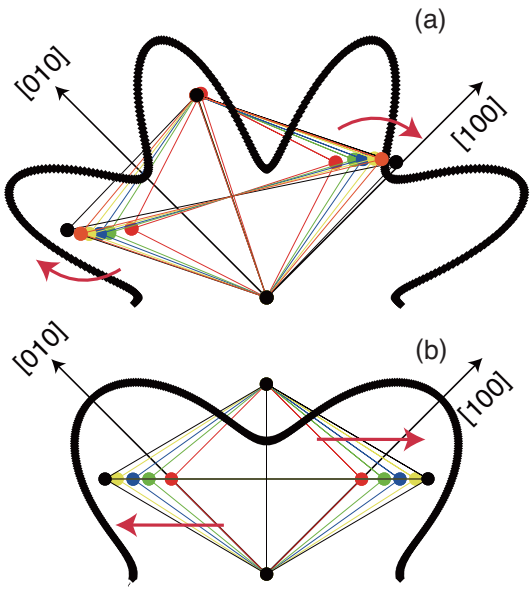} 
 \caption{ 
(Color online)  
FLL transformations as $T$ increases  for the two cases
where the bold curves are the Fermi velocity anisotropies:
(a) FLL changes from the square to the scalene for  the minimum model with $\beta_4=0.1$
and $\beta_8=0.3$. (b) FLL changes from the square to the isosceles 
for $\beta_4=0.1$ where the mirror symmetry is unbroken about the
maximum direction [100].
} 
\end{figure} 

It is important to notice that one of the two diagonal axes is kept to point along the 
direction of the velocity minimum during the deformation.
In other words, the base of the isosceles is always kept along the velocity
minimum direction.
The pairs of the vortices on the unit cell move along those symmetry
constraint axes, preserving the mirror symmetry.
This is contrasted with the present scalene case where the two 
diagonal axes rotate each other, thus no symmetry constraint during the deformation
because the mirror symmetry about the maximum direction is broken from the outset.
Comparing the two cases with either broken mirror symmetry or with unbroken symmetry,
it is concluded that a necessary condition, but not sufficient condition for
the scalene triangle FLL to occur is that in the anisotropic  velocity the
mirror symmetry is broken about the anisotropy maximum direction.
The broken mirror symmetry of the FLL that is real space 
is originated from the mirror symmetry breaking in the reciprocal space.
Furthermore, it is interesting to notice the experimental fact that
the appearance of the scalene FLL is tied with the appearance of the
tilted square lattice in Nb. As mentioned above, the square FLL is 
oriented along the velocity maximum direction at lower $T$,
that is a general theoretical fact within our model calculations.
Thus, in Nb the Fermi velocity maximum is situated along the 
direction tilted by 11$^{\circ}$ from the [001] axis
and the mirror symmetry about this axis must be broken.

In summary, we demonstrate that the first principles
band calculation combined with microscopic  Eilenberger analysis yields 
an excellent explanation of the observed FLL transformation in Nb under $H\parallel [001]$,
including the tilted square and scalene 
lattices without adjustable parameter.
This opens a door that this combined method provides a fundamental framework
to truly understand the vortex matter from first principles. 
We also show the possible reasons why the 
spontaneous mirror symmetry is broken in the scalene lattice
by constructing a minimum model for the Fermi velocity anisotropy.

We thank M. Laver and E. M. Forgan for stimulating discussions.
We also acknowledge useful discussion with  P. Miranovic.
This work is partly performed during a stay in Aspen Center for Physics.


\end{document}